\documentclass[a4paper,12pt]{article}

\newcommand{\sect}[1]{\setcounter{equation}{0}\section{#1}}

\begin{document}
\topmargin 0pt
\oddsidemargin 0mm

\renewcommand{\thefootnote}{\fnsymbol{footnote}}
\begin{titlepage}
\begin{flushright}
OU-HET 379\\
hep-th/0102113
\end{flushright}

\vspace{5mm}
\begin{center}
{\Large \bf Cardy-Verlinde Formula and AdS Black Holes}
\vspace{12mm}

{\large
Rong-Gen Cai\footnote{e-mail address: cai@het.phys.sci.osaka-u.ac.jp}
} \\
\vspace{8mm}
{\em Department of Physics, Osaka University,
Toyonaka, Osaka 560-0043, Japan} 
\end{center}
\vspace{5mm}
\centerline{{\bf{Abstract}}}
\vspace{5mm}
In a recent paper hep-th/0008140 by E. Verlinde, an interesting formula has 
been put forward, which relates the entropy of a conformal formal field in 
arbitrary dimensions to its total energy and Casimir energy. This formula has
been shown to hold for the conformal field theories that have AdS duals 
in the cases of AdS Schwarzschild black holes and AdS Kerr black holes. In
this paper we further check this formula with various black holes with AdS
asymptotics.  For the hyperbolic AdS black holes, the Cardy-Verlinde formula
is found to hold if we choose the ``massless'' black hole as the ground state,
 but in this case, the Casimir energy is negative. For the AdS 
Reissner-Nordstr\"om black holes in arbitrary dimensions and charged black 
holes in D=5, D=4, and D=7 maximally supersymmetric gauged supergravities,
the Cardy-Verlinde formula holds as well, but a proper internal energy which 
corresponds to the mass of supersymmetric backgrounds must be subtracted from 
the total energy. It is failed to rewrite the entropy of corresponding 
conformal field theories in terms of the Cardy-Verlinde formula for the AdS 
black holes in the Lovelock gravity.   

\end{titlepage}

\newpage
\renewcommand{\thefootnote}{\arabic{footnote}}
\setcounter{footnote}{0}
\setcounter{page}{2}

\sect{Introduction}

It is well-known that the entropy of a ($1+1)$-dimensional conformal field 
theory (CFT) can be given by the Cardy formula \cite{Card}
\begin{equation}
\label{1eq1}
S=2\pi \sqrt{\frac{c}{6}\left (L_0-\frac{c}{24}\right )},
\end{equation}
where $c$ is the central charge, $L_0$ denotes the product $ER$ of the 
total energy and radius of system, and the shift of $c/24$ is caused by the
Casimir effect, which is a finite volume effect.

In a recent paper by Erik Verlinde \cite{Verl}, it has been proposed that the
Cardy formula (\ref{1eq1}) can be generalized to the case in arbitrary 
dimensions.  Consider a conformal field theory living in $(1+n)$-dimensional 
spacetime described by the metric
\begin{equation}
\label{1eq2}
ds^2 = -dt^2 +R^2d\Omega_n^2,
\end{equation}
where $R$ is the radius of a $n$-dimensional sphere. The entropy of the CFT
can be given by the generalized Cardy formula (hereafter we refer to this as 
the Cardy-Verlinde formula),
\begin{equation}
\label{1eq3}
S= \frac{2\pi R}{\sqrt{ab}} \sqrt{E_c(2E-E_c)},
\end{equation}
where $E_c$ represents the Casimir energy,  $a$ and $b$ are two positive
coefficients which are independent of $R$ and $S$. For strong coupled
CFT's with AdS duals, the value of product $ab$ is fixed to $n^2$ exactly.
The above Cardy-Verlinde formula is then reduced to
\begin{equation}
\label{1eq4}
S=\frac{2\pi R}{n}\sqrt{E_c(2E-E_c)}.
\end{equation}
With given the total energy $E$ and the radius $R$, the Cardy-Verlinde
formula (\ref{1eq4}) gives the maximal entropy
\begin{equation}
S \le S_{\rm max}=\frac{2\pi R E}{n},
\end{equation}
when $E_c=E$. 
This is just the Bekenstein entropy bound \cite{Beke}\footnote{ The Bekenstein
entropy bound states that  the ratio of the entropy $S$ to the total energy
$E$ of a closed physical system with limited self-gravity, which fits in a 
sphere with radius $R$ in three spatial dimensions, obeys $S \le 2\pi R E$. 
In fact, the Bekenstein entropy bound is independent of  the spatial 
dimension. For a derivation of the Bekenstein bound in arbitrary dimensions 
see \cite{Bous}.}.   
  
In the spirit of  AdS/CFT correspondence \cite{Mald,Gubs,Witten1},
It was convincingly argued by Witten \cite{Witten2} that the thermodynamics
of a certain CFT at high temperature can be identified with the thermodynamics
of black holes in anti-de Sitter space (AdS). With this correspondence,
Verlinde checked the formula (\ref{1eq4}) using the thermodynamics of 
AdS Schwarzschild black holes in arbitrary dimensions and found it holds
exactly \cite{Verl}.  Furthermore the Cardy-Verlinde formula has been checked
more recently for the AdS Kerr black holes in \cite{Klem}, which corresponds
to a CFT residing in a rotating Einstein universe. Once again, this formula has 
been found to hold exactly. Some of recent works related to the entropy bound
and the Cardy-Verlinde formula are \cite{Kuta,Lin,Noji,Wang,Brus,Savo}.

No doubt, it is of great interest to prove the validity of the Cardy-Verlinde
formula for all CFT's. Of course, it might be a quite difficult task.  Having 
considered the absence of such a proof so far, it is worthwhile to do some 
further check for the Cardy-Verlinde 
formula in a larger extent than that in \cite{Verl} and \cite{Klem}, in 
order to see to what extent this formula is valid. This is just the aim of 
this paper.

In this paper  we choose some typical examples of black holes with AdS 
asymptotics to check the Cardy-Verlinde 
formula. In a $(n+2)$-dimensional AdS, except for the spherically symmetric 
AdS Schwarzschild black holes whose horizon is a $n$-dimensional sphere 
surface with positive constant curvature, There exist the so-called
hyperbolic AdS black holes whose horizon is a negative constant curvature
hypersurface. The thermodynamics of the latter is different from that of the
former. It would be interesting to see if the Cardy-Verlinde formula holds
in this case. This will be done in the next section. 

The gauged supergravity
can also be  realized as a self-consistent truncation of superstring or M
theory compactified on a compact manifold. The gauge group is the isometry
group of the compact manifold. In the course of AdS/CFT correspondence, the
gauge group is identified with the R-symmetry group of boundary CFT's. In this
sense the thermodynamics of AdS charged black holes can be viewed as that of
a certain CFT with a chemical potential. So we will check the Cardy-Verlinde 
formula in section 3 with AdS Reissner-Nordstr\"om black holes. There we will
also discuss the charged black holes in D=5, D=4 and D=7 maximally 
supersymmetric gauged supergravities. These theories can be regarded as the
self-consistent truncations of IIB supergravity on the $S^5$, 11-dimensional 
supergravity on the $S^7$ and $S^4$, respectively.

 In supergravity theories, higher derivative curvature terms occur as the
 corrections of the massive string states and string loop corrections in 
superstring theories. In the AdS/CFT correspondence, these corrections
 correspond to those of large $N$ expansion of boundary CFT's in the strong 
coupling limit. So it is also interesting to see if the Cardy-Verlinde formula
 still remains valid after including some of those corrections. However, it
 is in general quite difficult to find exact nontrivial
black hole solutions in higher derivative gravity theories, which are required
for exactly checking the Cardy-Verlinde formula. In section 4 we will consider
a special kind of Lovelock gravity theories, in which by choosing some special 
coefficients for each term in the action,  a simple, but exact  black hole 
solution can be found. Using this 
black hole solution, we examine the thermodynamics of corresponding CFT's.
We summarize our results in section 5 with brief discussions.


\sect{Hyperbolic AdS black holes in arbitrary dimensions }

In four dimensional spacetimes, it is believed generally that the horizon
of a black hole must be a sphere $S^2$, up to diffeomorphisms. However, it 
can be violated if the theory includes a negative cosmological constant. 
It was already found that except for the sphere case, namely, the horizon is 
a positive constant curvature hypersurface, black holes are allowed to exist 
with horizon which are zero or negative constant curvature hypersurface. In 
higher dimensional ($D\ge 4$) spaces, it is true as well.  For those so-called
topological black hole solutions in arbitrary dimensions see \cite{Birm}.

For a $(D=n+2)$-dimensional spherically symmetric black hole in AdS spacetime,
its thermodynamics corresponds to the one of a CFT living in 
$(n+1)$-dimensional
spacetime with topology $R \times S^n$. This case has been already checked
in \cite{Verl}. For black holes with zero curvature horizon, its 
thermodynamics is conformal invariant and the Casimir energy vanishes.
So the Cardy-Verlinde formula (\ref{1eq4}) is not applicable in this case.
As a result, in this section we discuss the AdS black holes with negative 
constant curvature horizon.  In this case, its thermodynamics corresponds
to the one of a CFT residing in a spacetime $R \times \Sigma_g^n$, where
$\Sigma_g^n$ denotes a $n$-dimensional negative constant curvature space, 
which can be a closed hypersurface with arbitrary high genus under 
appropriate identification.

The metric for a hyperbolic AdS black hole in a $(n+2)$-dimensional spacetime
can be written down as \cite{Birm}
\begin{equation}
\label{2eq1}
ds^2 = -f(r)dt^2 +f(r)^{-1}dr^2 +r^2 d\Sigma^2_n,
\end{equation}
where
\begin{equation}
f(r) =-1 -\frac{\omega_n M}{r^{n-1}} +\frac{r^2}{l^2}, \ \  
\omega_n=\frac{16\pi G}{n Vol(\Sigma_n)},
\end{equation}
$d\Sigma_n^2$ denotes the line element of a $n$-dimensional hypersurface
with constant curvature $-n(n+1)$,  $Vol(\Sigma_n)$ stands for the volume
of the hypersurface $\Sigma_n$, and $G$ is the Newtonian gravity constant.
This is a solution of Einstein equations with a negative cosmological 
constant $\Lambda = -n(n+1)/2l^2$. 
 
The solution (\ref{2eq1}) has some peculiar properties in the sense: 
(1) When the integration constant $M=0$, even though the solution
is locally an anti-de Sitter space, it has a black hole horizon $r_+=l$
with Hawking temperature $T_{\rm HK}$ and Bekenstein-Hawking entropy $S$, 
\begin{equation}
\label{2eq3}
T_{\rm HK}= \frac{1}{2\pi l}, \ \ \  S =\frac{l^n Vol(\Sigma_n)}{4G}.
\end{equation}
This is the so-called ``massless'' black hole;  (2) When $M >0$, the solution
(\ref{2eq1}) has only a black hole horizon satisfying
\begin{equation}
\label{2eq4}
M =\frac{r_+^{n-1}}{\omega_n}\left (\frac{r_+^2}{l^2}-1\right ),
\end{equation}
which implies that $r_+ >l$.  When $M <0$, however, it can have two black hole
horizons, which coincide as 
\begin{equation}
\label{2eq5}
M =M_{\rm ext}= -\left (\frac{2}{n+1}\right)
   \left (\frac{n-1}{n+1}\right)^{(n-1)/2} \frac{l^{n-1}}{\omega_n}.
\end{equation}
In this case, the coincident horizon $r_+^2=l^2(n-1)/(n+1)$, the Hawking
temperature vanishes, the black hole becomes an extremal one. It is the
peculiar property which causes the difficulty to choose an appropriate 
reference background in order to determine
the mass of hyperbolic black holes \cite{Birm,Emp}. In other words, there 
are some debates about the ground state of the hyperbolic AdS black holes.

Let us first suppose the ``massless'' black hole (\ref{2eq3}) as the ground 
state of the hyperbolic AdS black holes (\ref{2eq1}).  In this case, the 
constant $M$ is the mass of black holes, the temperature and entropy of black 
holes are 
\begin{eqnarray}
\label{2eq6}
&& T_{\rm HK} = \frac{1}{4\pi r_+}\left(\frac{(n+1) r_+^2}{l^2}-(n-1)\right),
    \nonumber \\
&& S =\frac{r_+^n Vol(\Sigma_n)} {4G}.
\end{eqnarray}
According to the prescription of AdS/CFT correspondence \cite{Gubs,Witten1},
the boundary spacetime in which the boundary CFT resides can be determined 
using the bulk metric, up to a conformal factor. It is due to the conformal 
factor that one can arbitrary rescale the boundary metric as one wishes.
In this paper, we rescale the boundary metric
so that the finite volume has a radius $R$ (this implies that $T> 1/R$ is 
assumed for the temperature $T$ of corresponding CFT's) \footnote{
In \cite{Verl}
the radius is taken to be the horizon radius of black holes.}. That is, the
boundary metric has the following form
\begin{equation}
\label{2eq7}
ds^2_b=\lim_{r \to \infty} \frac{R^2}{r^2}ds^2 =-\frac{R^2}{l^2}dt^2 +R^2 
   d\Sigma^2_n.
\end{equation} 
Thus the system has the finite volume $V= R^n Vol(\Sigma_n)$. 
The $(n+1)$-dimensional CFT corresponding to the hyperbolic  black holes 
has the energy $E$, temperature $T$ and entropy $S$ in the metric (\ref{1eq2}),
\begin{eqnarray}
\label{2eq8}
&& E =\frac{nl Vol(\Sigma_n)r_+^{n-1}}{16\pi GR}\left (\frac{r_+^2}{l^2}
    -1\right), \nonumber \\
&& T =\frac{l}{4\pi Rr_+}\left (\frac{(n+1)r_+^2}{l^2}-(n-1)\right), 
    \nonumber \\
&& S = \frac{r_+^n Vol(\Sigma_n)}{4G}.
\end{eqnarray}
Following \cite{Verl}, let us define the Casimir energy $E_c$ as
\begin{equation}
\label{2eq9}
E_c= n(E+pV -TS),
\end{equation}
where $p$ represents the pressure of CFT defined as $p=-\left(\frac{\partial E}
{\partial V}\right)_S$. With the help of (\ref{2eq8}), we obtain 
\begin{equation}
\label{2eq10}
E_c= -2 \frac{nlr_+^{n-1}Vol(\Sigma_n)}{16\pi GR}.
\end{equation}
Furthermore we have the extensive energy  
\begin{equation}
\label{2eq11}
2E-E_c =2\frac{n r_+^{n+1} Vol(\Sigma_n)}{16\pi GR l}.   
\end{equation}
With these quantities, we find that the entropy $S$ of CFT in (\ref{2eq8}) 
can be rewritten as
\begin{equation}
\label{2eq12}
S =\frac{2\pi R}{n}\sqrt{|E_c|(2E-E_c)}.
\end{equation}
Comparing with the Cardy-Verlinde formula (\ref{1eq4}), we find that indeed
the entropy of a CFT residing in a finite hyperbolic space can also be 
expressed in a form of the Cardy-Verlinde formula. But,
we find the Casimir energy is negative in this case! This feature can be 
traced back to the peculiar properties of hyperbolic AdS black holes 
discussed above. In other words, it is related to the existence of the
 ``massless'' black holes and ``negative mass'' black holes.

One may wonder if the difficulty of negative Casimir energy can be 
circumvented through choosing the extremal black hole (\ref{2eq5}) as
the ground state of black holes. In that case, the thermodynamics of black 
holes is still given by (\ref{2eq6}), but the mass of black holes becomes
$M-M_{\rm ext}$. After a simple repeat as the above, one has to be led to 
the conclusion that the entropy cannot be expressed in terms of the 
Cardy-Verlinde formula in this case.

\sect{Charged AdS black holes }

\subsection{AdS Reissner-Nordstr\"om black holes in arbitrary dimensions}

The metric of a $(n+2)$-dimensional AdS Reissner-Nordstr\"om black hole
is \cite{CS,Cham}
\begin{equation}
\label{3eq1}
ds^2= -f(r)dt^2 +f(r)^{-1}dr^2 +r^2 d\Omega_n^2,
\end{equation}
where $d\Omega_n^2$ denotes the line element of a unit $n$-dimensional 
sphere and the function $f$ is given by
\begin{equation}
\label{3eq2}
f(r)= 1-\frac{m}{r^{n-1}} +\frac{\tilde q^2}{r^{2n-2}} +\frac{r^2}{l^2}. 
\end{equation}
When $m/2 =|\tilde q| $, this solution is supersymmetric and the function $f$ 
has the form 
\begin{equation}
\label{req3}
f(r)= \left( 1-\frac{m}{2r^{n-1}}\right)^2 +\frac{r^2}{l^2}.
\end{equation}
Obviously, in this case, this solution does not represent a black hole and the
singularity at $r=0$ becomes naked. 

For the convenience of discussions below, let us rewrite the solution 
(\ref{3eq1}) in terms of ``isotropic'' coordinates. Defining 
\begin{equation}
m = \mu + 2q, \ \ \  \tilde q^2 =q(\mu +q), \ \ \ r^{n-1} \to r^{n-1}+q,
\end{equation}
we can change (\ref{3eq1}) to the following form   
\begin{equation}
\label{req5}
ds^2 = H^{-2} f(r) dt^2 + H^{2/(n-1)}(f(r)^{-1}dr^2 +r^2 d\Omega_n^2),
\end{equation}
where 
\begin{equation}
f(r) = 1-\frac{\mu}{r^{n-1}} +\frac{r^2}{l^2}H^{2n/(n-1)}, \ \ \
  H=1+\frac{q}{r^{n-1}}.
\end{equation}
In this coordinates, the supersymmetric solution corresponds to the case 
when $\mu =0$.  The horizon $r_+$ of black hole is determined by the equation
\begin{equation}
 \mu = r_+^{n-1} +\frac{\rho ^{2n/(n-1)}}{l^2 r_+^{n-1}}, \ \ \ 
 \rho =r_+^{n-1} +q.
\end{equation}

To find the thermodynamic quantities of black hole is straightforward. The 
mass $M$, entropy $S$ and Hawking temperature $T_{\rm HK}$ are
\begin{eqnarray}
\label{req8}
&& M =\frac{n Vol(S^n)}{16\pi G}\left (\mu +2 q \right), \nonumber \\
&& S =\frac{Vol(S^n)}{4G}\rho ^{n/(n-1)}, \nonumber \\
&& T_{\rm HK}= \frac{r_+^{n-1}}{4\pi \rho^{n/(n-1)}}\left ( (n-1)
   +\frac{2n \rho^{2n/(n-1)}}{l^2 r_+^{n-1}} -\frac{(n-1)\rho^{2n/(n-1)}}
   {l^2 r_+^{2n-2}}\right), 
\end{eqnarray}
respectively, where $Vol(S^n)$ stands for the volume of the unit 
$n$-dimensional sphere. The chemical potential $\phi$ associated with the 
physical electric charge $\tilde q$ is 
\begin{equation}
\phi =\frac{n Vol(S^n)}{16\pi G}\frac{2\tilde q}{\rho}.
\end{equation}
As expected, these thermodynamic quantities satisfy the first law of black
hole thermodynamics
\begin{equation}
dM = T_{\rm HK} dS +\phi d\tilde q.
\end{equation}

Rescaling the boundary metric so that  the $n$-dimensional sphere has the 
radius $R$ and the volume $V=R^nVol(S^n)$, in the spirit of AdS/CFT 
correspondence, we have the energy,
temperature, and chemical potential of the corresponding CFT in the metric
(\ref{1eq2}), 
\begin{eqnarray}
\label{3eq7}
E = \frac{l}{R} M, \ \  T=\frac{l}{R}T_{\rm HK}, \ \ \Phi=\frac{l}{R}\phi,
\end{eqnarray}
respectively. The entropy and electric charge of CFT are still given by $S$
and $\tilde q$. From (\ref{req8}), we can see that the energy of CFT can 
be separated to two parts: one of them is proportional to $q$, which is the
contribution of supersymmetric background; the other is proportional to 
$\mu$, which corresponds to the contribution of thermodynamic excitations. 
Let us define
\begin{equation}
\label{req12}
E_q =\frac{nl Vol(S^n)}{16\pi GR}\cdot 2q.
\end{equation}
Reasonably, we can view the proper internal energy $E_q$ as the zero 
temperature energy of CFT, which makes the contribution to the free energy,
but not to the entropy.  Following \cite{Verl}, we define the Casimir energy 
in this case as
\begin{equation}
\label{3eq8}
E_c= n(E+pV -TS -\Phi \tilde q),
\end{equation}  
where the pressure $p$ is defined as $p=-\left(\frac{\partial E}{\partial V}
\right)_{S,\tilde q}$. Here it is worth stressing that when the electric charge
vanishes, we have the pressure $p= E/nV$ because  we are considering CFT's.
When the electric charge does not vanish, however,  the proper internal
energy (zero temperature energy) $E_q$ does not make its contribution to the
pressure. Therefore the pressure should have the following form
\begin{equation}
\label{req14}
p = \frac{E-E_q}{n V}.
\end{equation}
With this relation, substituting those quantities in (\ref{3eq7}) into 
(\ref{3eq8}) yields
\begin{equation}
\label{3eq9}
E_c=2\frac{nl Vol(S^n)r_+^{n-1}}{16\pi GR}.
\end{equation}
Using this Casimir energy, it is easy to find that the entropy of corresponding
CFT's for the AdS Reissner-Nordstr\"om black holes can be expressed as
\begin{equation}
\label{3eq11}
S=\frac{2\pi R}{n}\sqrt{E_c(2(E-E_q)-E_c)},
\end{equation}
where $E_q$ is the proper internal energy, given by (\ref{req12}). 
Here the difference from the standard Cardy-Verlinde formula (\ref{1eq4}) 
is the emergence of the proper internal energy $E_q$, which must be subtracted 
from the total energy. This result is  reasonable because the proper internal
energy (zero temperature energy) does not make its contribution to the entropy
of CFT.  In fact, following \cite{Verl}, we can also ``derive'' the formula 
(\ref{3eq11}) after considering there is an additional zero temperature 
energy in a certain thermodynamic system. The formula (\ref{3eq11}) is 
encouraging and the 
observation (\ref{req14})
is also interesting.  To see whether it is universal, in the following 
subsections we will check the formula (\ref{3eq11}) with the charged black 
holes in the maximally supersymmetric gauged supergravities, in which some 
scalar fields are present.

\subsection{Charged black holes in D=5 gauged supergravity}

In this subsection we discuss the case of black holes in D=5, N=8 gauged
supergravity. This solution has been found in \cite{Behr} (also see  
\cite{Cvetic}) as a special case (STU model) in the D=5, N=2 gauged 
supergravity.

The black hole solution has the metric
\begin{equation}
\label{3eq12}
ds^2 =-(H_1H_2H_3)^{-2/3} fdt^2 +(H_1H_2H_3)^{1/3}(f^{-1}dr^2 +r^2d\Omega_3^2),
\end{equation}
where 
\begin{equation}
\label{3eq13}
f =1 -\frac{\mu}{r^2}+r^2l^{-2} H_1 H_2 H_3, \ \ \  H_i=1+\frac{q_i}{r^2}, \ \ 
 i =1, 2, 3.
\end{equation}
There are three real scalar fields $X^i$ and three gauge potentials $A^i$  
\begin{equation}
\label{3eq14}
X^i = H_i^{-1}(H_1H_2 H_3)^{1/3}, \ \ \ A^i_t = \frac{\tilde q_i}{r^2 +q_i},
\ \ i=1, 2, 3.
\end{equation}
Here the charges $q_i$ are related to the physical electric charges 
$\tilde q_i$ via
\begin{equation}
\label{3eq15}
q_i=\mu \sinh^2\beta_i, \ \ \ \tilde q_i =\mu \sinh\beta_i \cosh\beta_i.
\end{equation}
The solution (\ref{3eq12}) has black hole horizon $r_+$ obeying the following
equation
\begin{equation}
\label{3eq16}
\mu = r_+^2 \left (1+\frac{1}{l^2 r_+^4} \prod _{i=1}^3 \rho_i \right), \ \
 \rho_i =r_+^2 +q_i.
\end{equation}  
The mass, Hawking temperature and entropy of black holes are 
\begin{eqnarray}
\label{3eq17}
&&M =\frac{\pi}{4G}\left (\frac{3}{2}\mu +\sum_{i}q_i\right), \nonumber \\
&& T_{\rm HK} =\frac{r_+^2}{2\pi \sqrt{\prod_i \rho_i}}\left [
   1-\frac{\prod_i \rho_i}{l^2 r_+^4}\left (1-r_+^2 \sum_i \frac{1}{\rho_i}
   \right) \right], \nonumber \\
&& S =\frac{\pi^2}{2G}\sqrt{\prod_i \rho_i}.
\end{eqnarray}
The associated chemical potentials with the electric charges $\tilde q_i$ are
\begin{equation}
\phi_i = \frac{\pi}{4G}\frac{\tilde q_i}{\rho_i}.
\end{equation}
As required, the first law of black hole thermodynamics is satisfied: 
\begin{equation}
dM = T_{\rm HK}dS +\sum_i \phi_i d\tilde q_i.
\end{equation}
Rescaling the boundary metric so that the three-dimensional sphere
has the radius $R$, and using the relations (\ref{3eq8}) and (\ref{req14})
 we obtain the Casimir energy
\begin{equation}
\label{3eq20}
E_c =\frac{\pi l}{4GR}(\sum_i \rho_i-\sum_iq_i) = \frac{3\pi lr_+^2}{4GR}.
\end{equation}
With this Casimir energy,  the entropy of 
corresponding  CFT can be written in the following form
\begin{equation}
\label{3eq21}
S= \frac{2\pi R}{3}
   \sqrt{E_c(2(E-E_q)-E_c)}, 
\end{equation}
where the proper internal energy $E_q$ and the thermal excitation energy are 
\begin{equation}
E_q =\frac{\pi l}{4GR}\sum_i q_i, \ \ \ E-E_q=\frac{3\pi l}{8GR}\left (r_+^2 
  +\frac{1}{l^2 r_+^2}\prod_i \rho_i\right).
\end{equation}
Clearly, the expression (\ref{3eq21}) is a special case of (\ref{3eq11}) when
$n=3$, although the thermodynamics of the black hole solutions (\ref{3eq12}) is
different from that of D=5  AdS Reissner-Nordstr\"om black holes 
because of the presence of three real scalar fields. Of course, the former
degenerates to  the latter when three charges are equal to each other. This
can be seen from the solution (\ref{3eq12}).

\subsection{Charged black holes in D=4 gauged supergravity}

The black hole solution in D=4, N=8 gauged supergravity has been found in
\cite{Duff}. The metric has the form
\begin{equation}
\label{3eq22}
ds^2 = -(H_1H_2H_3H_4)^{-1/2} fdt^2 +(H_1H_2H_3H_4)^{1/2}(f^{-1}dr^2
   +r^2 d\Omega_2^2),
\end{equation}
where
\begin{equation}
f = 1-\frac{\mu}{r} +l^{-2}r^2 \prod_{i=1}^4H_i,  \ \  H_i =1+\frac{q_i}{r},
\ \ \ i=1, 2, 3, 4.
\end{equation}
The four electric potentials are
\begin{equation}
A^i _t = \frac{\tilde q_i}{r +q_i}, \ \ \  i=1, 2, 3, 4.
\end{equation}
The physical charges $\tilde q_i$ are related to the charges $q_i$ as the form
(\ref{3eq15}). For the black hole solution (\ref{3eq22}), one has the
horizon $r_+$ which satisfies the following equation,
\begin{equation}
\mu = r_+\left( 1+\frac{1}{l^2 r_+^2} \prod_i \rho_i \right), \ \ \ \rho_i =
  r_+ +q_i, \ \ \ i=1, 2, 3, 4.
\end{equation}
The thermodynamics associated with black hole horizon can be easily found.
The mass, Hawking temperature and entropy are
\begin{eqnarray}
&& M = \frac{1}{4G}(2\mu + \sum_i q_i), \nonumber \\
&& T_{\rm HK} =\frac{r_+}{4\pi \sqrt{\prod_i \rho_i}}\left ( 1-\frac{\prod_i 
   \rho_i}{l^2 r_+^2}+ \frac{\prod_i \rho_i}{l^2 r_+}\sum_i \frac{1}{\rho_i}
    \right), \nonumber \\
&& S =\frac{\pi}{G}\sqrt{\prod_i \rho_i}, 
\end{eqnarray}
respectively.  The chemical potentials $\phi_i$ conjugating to the 
charges $\tilde q_i$ are 
\begin{equation}
\phi _i = \frac{1}{4G}\frac{\tilde q_i}{\rho_i}, \ \ \  i=1, 2, 3, 4.
\end{equation}
Once again, rescaling the boundary metric so that the two-dimensional
sphere has the radius $R$, and repeating the calculations as the
previous subsection, one has the Casimir energy
\begin{equation}
E_c= \frac{l}{4 GR}( \sum_i \rho_i-\sum_i q_i) =\frac{lr_+}{GR}.
\end{equation}
And the entropy can be rewritten as 
\begin{equation}
\label{3eq29}
S =\frac{2\pi R}{2}
   \sqrt{E_c(2(E-E_q)-E_c)}, 
\end{equation}
where the proper internal energy and the thermal excitation energy are
\begin{equation}
E_q=\frac{l}{4GR}\sum_iq_i, \ \ \  E-E_q=\frac{l}{2GR}\left(r_+ 
     +\frac{1}{l^2r_+} \prod_i \rho_i\right).
\end{equation} 
The expression of entropy is the case of D=4 AdS Reissner-Nordstr\"om
black holes.  Although the thermodynamics of the solution (\ref{3eq22}) is
also different from the one of D=4 AdS Reissner-Nordstr\"om black 
holes, the entropy of corresponding CFT's falls into the Cardy-Verlinde 
formula, which indicates the universality of the Cardy-Verlinde formula.

\subsection{Charged black holes in D=7 gauged supergravity}

The black hole solution in the D=7, N=4 gauged supergravity has the
form \cite{Liu,Cvetic}
\begin{equation}
\label{3eq30}
ds^2 = -(H_1H_2)^{-4/5} fdt^2 +(H_1H_2)^{1/5}(f^{-1}dr^2 +r^2d\Omega_5^2),
\end{equation}
where
\begin{equation}
f(r)= 1-\frac{\mu}{r^4} +r^2l^{-2}H_1 H_2, \ \ \  H_i =1+\frac{q_i}{r^4},
 \ \  i=1, 2. 
\end{equation}
The two gauge potentials in the solution (\ref{3eq30}) are
\begin{equation}
A^i_t = \frac{\tilde q_i}{r^4+q_i},  \ \ \ i=1, 2.
\end{equation}
As the case of D=5 or D=4, the physical charges $\tilde q_i $ are also 
related to
the charges $q_i$ via the relation (\ref{3eq15}). A standard calculation 
gives the thermodynamics of black hole solution (\ref{3eq30}):
\begin{eqnarray}
&& M = \frac{\pi^2}{4G}\left (\frac{5}{4}\mu +\sum_i q_i\right), \nonumber \\
&& T_{\rm HK}= \frac{r_+^3}{\pi \sqrt{\prod_i \rho_i}}
  \left (1-\frac{\prod_i \rho_i}{2r_+^6 l^2} +\frac{\prod_i \rho_i}{r_+^2 l^2}
   \sum_i \frac{1}{\rho_i}\right ), \nonumber \\
&& S= \frac{\pi^3 r_+}{4G} \sqrt{\prod_i \rho_i}, \nonumber \\
&& \phi_i = \frac{\pi^2}{4G}\frac{\tilde q_i}{\rho_i},
\end{eqnarray}
where the constant $\mu$ is related to the black hole horizon $r_+$ as
\begin{equation}
\mu = r_+^4 +\frac{1}{r_+^2 l^2}\prod_i \rho_i,
 \ \ \ \rho_i = r_+^4 +q_i, \ \ \  i=1, 2.
\end{equation}
The first law  here $dM = T_{\rm HK}dS +\sum_i \phi_i d\tilde q_i$ can be 
easily checked.    

With the relations (\ref{3eq8}) and (\ref{req14}), we find the Casimir energy 
in this case is
\begin{equation}
E_c =\frac{5\pi^2 lr_+^4}{8GR},
\end{equation}
and the entropy of corresponding CFT has the form
\begin{equation}
\label{3eq36}
S=\frac{2\pi R}{5} \sqrt{E_c(2(E-E_q)-E_c)},
\end{equation}
where the proper internal energy and the thermal excitation  energy are
\begin{equation}
E_q =\frac{\pi^2 l}{4GR}\sum_i q_i, \ \ \  E-E_q=\frac{5\pi^2 l}{16 GR}
  \left( r_+^4 +\frac{1}{r_+^2 l^2}\prod_i \rho_i \right ).
\end{equation}
Once again, the entropy of corresponding CFT's to the charged black holes
in D=7 gauged supergravity has the form of Cardy-Verlinde formula. 
Note that the solution (\ref{3eq30}) does not go to the one for a D=7 AdS 
Reissner-Nordstr\"om black hole even when two charges are equal, $q_1=q_2$. 
This example further manifests that the Cardy-Verlinde formula (\ref{3eq11})
and the observation (\ref{req14}) on the pressure are universally valid
for charged AdS black holes.


\sect{AdS black holes in higher derivative gravity}

In this section we consider the AdS black holes in a special class of
Lovelock gravity, which may be regarded as the most general generalization
to higher dimensions of Einstein gravity. The Lovelock action is a sum of
 the dimensionally continued Euler charactertics of all dimensions below the
spacetime dimension ($D$) under consideration. The Lovelock action has an 
advantage that the resulting equations of motion contain no more than 
second derivatives of metric, as the pure 
Einstein-Hilbert action, but it includes $[D/2]$ arbitrary coefficients, 
which make it difficult to extract physical information from the solutions of 
equations of motion. It is possible to reduce those coefficients to 
two ones: a cosmological constant and a gravitational constant. By embedding 
the Lorentz group $SO(D-1,1)$ into a larger group, the anti-de Sitter 
group $SO(D-1,2)$, the Lovelock theory is divided into two different branches 
according to the spacetime dimension: odd dimensions and even 
dimensions. In odd dimensions, the action is the Chern-Simons form for the 
anti-de Sitter group; in even dimensions, it is the Euler density constructed
with the Lorentz part of the anti-de Sitter curvature tensor. For details 
see \cite{Bano}.

The metric of a $(D=n+2)$-dimensional AdS black holes in the dimensionally
continued gravity theory is \cite{CS,Bano}
\begin{equation}
\label{4eq1}
ds^2 =- f(r) dt^2 +f(r)^{-1} dr^2 + r^2 d\sigma^2_n,
\end{equation}
where
\begin{equation}
f(r) = \left \{
\begin{array}{ll}
 k-\left(\frac{2M}{r}\right)^{\frac{1}{m-1}} +\frac{r^2}{l^2} & {\rm for}\ \
   D=2m, \\
 k-M^{\frac{1}{m-1}} +\frac{r^2}{l^2} & {\rm for}\ \ D =2m -1,
\end{array} \right.
\end{equation}
where $M$ is an integration constant and can be explained as 
the mass of black holes, in this case, it is implied that the anti-de Sitter
space is viewed as the ground state of black holes \cite{CS}. 
$d\sigma_n^2=\gamma_{ij}(x)dx^idx^j$
denotes the line element of a $n$-dimensional hypersurface with constant
curvature $n(n-1)k$. Without loss of generality, one may take $k=1$, $0$ and 
$-1$, respectively. When $k=1$ the hypersurface $\sigma_n$ is a positive 
constant curvature space, a simple case is just $n$-dimensional unit 
sphere $S^n$, as discussed in the above. When $k=-1$, the hypersurface is a 
negative constant curvature space. In this case, one can construct a closed 
hypersurface with arbitrary high genus via appropriate identification. 
When $k=0$, the hypersurface is a zero curvature space, because of the reason
explained in Section 2, we will not discuss this case.

In the solution (\ref{4eq1}), the horizon $r_+$ is determined by the equation
\begin{equation}
\label{4eq3}
M =\left \{ 
\begin{array}{ll}
\frac{r_+}{2}\left(k +\frac{r_+^2}{l^2}\right )^{m-1} & {\rm for }\ \  D=2m, 
     \\ 
\left( k+\frac{r_+^2}{l^2} \right)^{m-1} & {\rm for}\ \  D=2m-1.
\end{array} \right.
\end{equation}
The Hawking temperature of black holes can be easily calculated, which is
\begin{equation}
T_{\rm HK} = \left \{
\begin{array}{ll}
\frac{1}{4\pi (m-1) r_+}\left (k +\frac{(2m-1)r_+^2}{l^2}\right) & {\rm for}
  \ \  D=2m, \\
\frac{r_+}{2\pi l^2} & {\rm for}\ \ D=2m-1.
\end{array} \right.
\end{equation}
For the black holes in higher derivative gravity theories, the entropy is
not simply one quarter of horizon area.   In \cite{CS} we have presented
a method to obtain the entropy of black holes in higher derivative gravity
theories. That method is based on the fact that all black holes must obey 
the first law of thermodynamics $dM =T_{\rm HK}dS +\cdots$. Integrating the 
first law, we have
 \begin{equation}
S =\int T_{\rm HK}^{-1} dM =\int^{r_+}_0 T^{-1}_{\rm HK} \left(
    \frac{\partial M} {\partial r_+}\right) dr_+,
\end{equation}
where we have imposed the physical assumption that the entropy vanishes when
the horizon of black holes shrinks to zero. Evidently the entropy gained in
 this
way is independent of the choice of ground state of black holes. With this 
formula,  we can obtain easily the entropy of black holes in (\ref{4eq1})
\begin{equation}
\label{4eq6}
S =\left \{
\begin{array}{ll}
 \pi l^2\left ( \left(k+\frac{r_+^2}{l^2}\right)^{m-1}-k\right) &
   {\rm for}\ \  D=2m, \\
4\pi (n-1) l \sum_{i=0}^{m-2}\left(
\begin{array}{c}
m-2 \\ i 
\end{array} \right) 
\frac{1}{2i +1}\left(\frac{r_+}{l}\right)^{2i+1}k^{m-2-i}
   & {\rm for}\ \ D=2m-1.
\end{array} \right.
\end{equation}

Now we are ready to check the Cardy-Verlinde formula with the AdS black holes
in higher derivative gravity. Rescaling the metric so that the constant 
curvature hypersurface has the radius $R$, we then have the energy $E$ and
temperature $T$ of the corresponding CFT's 
\begin{equation}
\label{4eq7}
 E = \frac{l}{R}M, \ \ \  T=\frac{l}{R}T_{\rm HK},
\end{equation}
and the entropy $S$ of the CFT's is still given by the entropy $(\ref{4eq6})$
of black holes. 
 
In the case for the even dimensional black holes, namely, $D=2m$, we find 
the Casimir energy is
\begin{equation}
\label{4eq8}
E_c= \frac{kr_+l}{2R}\left [ 2m-1 -\frac{l^2}{r_+^2}\left(\left(k+\frac{r_+^2}
      {l^2}\right)^{m-1}-k\right)\right],
\end{equation}
and furthermore we have
\begin{equation}
\label{4eq9}
2E-E_c=\frac{r_+l}{2R}\left[ \left(2-\frac{l^2}{r_+^2}k\right) 
     \left(k+\frac{r_+^2}{l^2}\right)^{m-1}- (2m+1)k 
   +\frac{l^2}{r_+^2}k^2\right].
\end{equation}
So for both cases $k=\pm 1$, we cannot put the entropy in (\ref{4eq6})
into the form of Cardy-Verlinde formula.

In the case for odd dimensions ($D=2m-1$), the Casimir energy is
\begin{equation}
\label{4eq10}
E_c= \frac{2(m-1)l}{R}\left(k+\frac{r_+^2}{l^2}\right)^{m-1}
    -\frac{2(m-1)(2m-3)r_+}{R}
 \sum_{i=0}^{m-2}\left(
\begin{array}{c}
m-2 \\ i 
\end{array} \right) 
\frac{1}{2i +1}\left(\frac{r_+}{l}\right)^{2i+1}k^{m-2-i}.
\end{equation}
Again, we cannot put the entropy in (\ref{4eq6}) into the form 
of the Cardy-Verlinde formula. To clearly see this, let us consider a special
dimension $D=5$. In this case, the action of the gravity theory is the 
Einstein-Hilbert action plus a Gauss-Bonnet term. For such a black hole,
the entropy is
\begin{equation}
\label{4eq11}
S=8\pi r_+ \left(k +\frac{r_+^2}{3l^2}\right).
\end{equation}
And the Casimir energy for the corresponding CFT is found to be
\begin{equation}
\label{4eq12}
E_c= \frac{4kl}{R}\left (k-\frac{r_+^2}{l^2} \right),   
\end{equation}
and the extensive energy is
\begin{equation}
2E-E_c = \frac{2l}{R}\left(\left(2k +\frac{r_+^2}{l^2}\right)^2 -5k^2\right).
\end{equation}
This special example clearly indicates that the entropy (\ref{4eq11}) does
not fall into the form of Cardy-Verlinde formula (\ref{1eq4}).

\sect{Conclusions}

The Cardy-Verlinde formula recently proposed by E. 
Verlinde \cite{Verl}, relates the entropy of a  certain CFT to its 
total energy and Casimir energy in arbitrary dimensions. In the spirit
of AdS/CFT correspondence, this formula has been shown to hold exactly
for the cases of AdS Schwarzschild black holes and AdS Kerr black holes. 

In this paper we have further checked the Cardy-Verlinde formula with some
typical examples of black holes with AdS asymptotics. They are hyperbolic  
AdS black holes, AdS Reissner-Nordstr\"om black holes, charged black holes
in D=5, D=4, and D=7 maximally supersymmetric gauged supergravities, and 
AdS black holes in higher derivative gravity.  For the hyperbolic AdS black 
holes, the formula holds if we choose the ``massless'' black hole as the 
ground state of black holes (otherwise, this formula will no longer hold),
but in this case, the Casimir energy is found to be negative 
[see (\ref{2eq10})]!  Obviously, further investigations are needed for the
hyperbolic AdS black holes. In fact, the understanding of the AdS/CFT 
correspondence is poor  for the thermodynamics of the hyperbolic black 
holes so far \cite{Emp}.

 For the AdS
Reissner-Nordstr\"om black holes in arbitrary dimensions and charged black 
holes in D=5, D=4 and D=7 maximally supersymmetric gauged supergravities, 
the Cardy-Verlinde formula can also hold by subtracting the proper internal
energy  from the total energy [see (\ref{3eq11})].  The proper internal energy
corresponds to the contribution of supersymmetric backgrounds. In the 
thermodynamics of corresponding CFT's, we can view the proper internal
energy as the zero temperature energy, which has the contribution to the 
free energy, but not to the entropy of thermodynamic system. Therefore our 
result (\ref{3eq11}) is reasonable and can be viewed as an extension of 
Cardy-Verlinde formula (\ref{1eq4}).  In addition, it might be worth 
mentioning that for the corresponding CFT to the charged AdS
black holes, its pressure is given by (\ref{req14}), namely $p=(E-E_q)/nV$.
The quantity $E-E_q$ has an interpretation as the thermal excitation
energy of CFT's.  

We have also considered  the AdS black holes in the Lovelock gravity and found
that the entropy of corresponding CFT's cannot be put into the Cardy-Verlinde
formula. This seems reasonable since the Cardy-Verlinde formula (\ref{1eq3})
was derived through the assumption that the first subleading correction
to the extensive part of the energy scales like $E_c(\lambda S,\lambda V)
=\lambda ^{1-2/n}E_c(S,V)$. The corrections from higher derivative terms
are beyond the scope of the scaling.

\section*{Acknowledgments}

The author would like to thank N. Ohta for discussions. 
The work  was supported in part by the Japan Society for the 
Promotion of Science and in part by Grants-in-Aid for Scientific Research
Nos. 99020, 12640270 and in part by Grant-in-Aid on the Priority Area:
Supersymmetry and Unified Theory of Elementary particles.


\end{document}